\documentclass[12pt]{article}

\usepackage[margin=1in]{geometry}
\usepackage{graphicx}
\usepackage{amsmath}
\usepackage{amssymb}
\usepackage{setspace}
\usepackage[colorlinks=true,linkcolor=blue,citecolor=blue,urlcolor=blue]{hyperref}
\usepackage{booktabs}
\usepackage{caption}
\usepackage{subcaption}
\usepackage[numbers,sort&compress]{natbib}
\usepackage{float}
\usepackage{microtype}

\begin{document}

%% ============================================================
%% TITLE
%% ============================================================
\begin{center}
{\Large\textbf{Characterization of Photopolymerized Microscopic Chiral
Structures Using Photonic Orbital Angular Momentum}}
\end{center}

\begin{center}
Jing Xu,$^{1}$ Rik Strobbe,$^{1}$ Yovan de Coene,$^{1}$ Renaud A.~L.~Vall\'ee,$^{2,*}$ and Koen Clays$^{1,*}$
\end{center}

\begin{flushleft}
\small
$^{1}$Molecular Imaging and Photonics, Department of Chemistry, KU Leuven, Leuven, Belgium\\
$^{2}$Univ.\ Bordeaux, CNRS, CRPP, UMR~5031, F-33600 Pessac, France\\[0.3em]
$^{*}$Corresponding authors: \href{mailto:renaud.vallee@u-bordeaux.fr}{renaud.vallee@u-bordeaux.fr}; \href{mailto:koen.clays@kuleuven.be}{koen.clays@kuleuven.be}\\[0.3em]
\end{flushleft}

\vspace{1em}

%% ============================================================
%% ABSTRACT
%% ============================================================
\noindent\textbf{Abstract}

\noindent
The controlled fabrication and chiroptical characterization of microscale
chiral structures remain central challenges in photonics, sensing, and
metamaterial engineering. Here we demonstrate an accessible, low-cost
platform that combines digital micromirror device-enabled maskless photolithography with
capillarity-induced self-assembly to produce polymer chiral
microstructures of deterministic handedness, and a liquid-crystal spatial
light modulator to generate vortex beams for their
characterization via helical dichroism (HD). Using a standard 532~nm
laser, we observe HD signals of approximately 30\% for microstructures
with a characteristic diameter of $\sim$15~$\mu$m. Rigorous
finite-difference time-domain simulations performed on
three-dimensional geometries reconstructed from high-resolution Scanning Electron Microscopy  data
reproduce the experimental HD spectra and confirm the role of structural
handedness in driving the differential orbital angular momentum (OAM)
response. Near-mirror-symmetric HD spectra for opposite-handed
enantiomers, combined with a vanishing response for achiral controls,
establish OAM as a robust and spatially selective chiral probe at the
microscale. Crucially, both fabrication and characterization rely on
equipment standard in an optics laboratory, without recourse to
femtosecond sources, plasmonic substrates, or costly photoresists. These
results open practical pathways toward OAM-driven chiral sensing,
enantioselective detection, and photonic logic devices.

\vspace{1em}

\newpage

%% ============================================================
\section{Introduction}
%% ============================================================

Chirality---the geometric property by which an object cannot be
superimposed onto its mirror image---pervades nature from the molecular
to the macroscopic scale~\cite{Qu2000,Barron2008,Blackmond2010}.
Left- and right-handed enantiomers govern biological processes at the
molecular level while also manifesting in macroscopic forms such as
mollusk shells, plant tendrils, and spiral
galaxies~\cite{Jiang2019,Godinho2010,Wang2013}. The ability to
discriminate and characterize chirality is of critical importance in
pharmaceuticals~\cite{PanDrugs2025,Patel2008}, biochemical
research~\cite{Xu2023,Li2021}, food safety~\cite{Wang2023food}, and
environmental monitoring~\cite{Wang2023food,Garcia2024}, where accurate enantiomer
identification is essential. Beyond molecular systems, advances in
nanophotonics have enabled the fabrication of artificial chiral
metamaterials whose optical performance---including polarization control
and emission properties---depends critically on precisely defined
geometric chirality~\cite{Zhao2017}.

Conventional chiroptical techniques such as optical rotation and circular
dichroism (CD) rely on photonic spin angular momentum
(SAM)~\cite{Gottarelli2008,Govorov2011,Fan2010}. Their signals arise from
interference between dominant electric-dipole (E1) transitions and much
weaker magnetic-dipole (M1) or electric-quadrupole (E2) transitions,
making them intrinsically faint. They typically require long optical
paths or large sample volumes and, critically, their sensitivity drops
sharply when structural feature sizes greatly exceed the probing
wavelength, rendering them poorly suited for thin films and microscale
chiral objects~\cite{Barron2009,Rodger2021,Choi2019}.

Photonic orbital angular momentum (OAM), characterized by a helical phase
front and an unbounded topological charge $\ell$, offers a fundamentally
different interaction pathway~\cite{Krenn2017}. OAM-based chiroptical
interactions are dominated by E1--E2 coupling, which can be orders of
magnitude stronger than the E1--M1 channel limiting
CD~\cite{Forbes2018}. Furthermore, the transverse ring diameter of a
vortex beam scales with $|\ell|$, enabling spatial tuning to match
microscale structural features. A vortex beam can be written as

\begin{equation}
    E(r,\varphi) = A(r)\,e^{i\ell\phi},
    \label{eq:vortex}
\end{equation}

\noindent where $A(r)$ is the radial amplitude envelope, $\phi$ is the
azimuthal angle, and $\ell$ is the topological charge whose sign
determines the handedness of the helical wavefront~\cite{Krenn2017}.
Helical dichroism (HD), the OAM analogue of CD, quantifies the
differential optical response of a chiral object illuminated by vortex
beams of opposite topological charges:

\begin{equation}
    \mathrm{HD(\ell)} = 2\,\frac{R_{+\ell} - R_{-\ell}}{R_{+\ell} + R_{-\ell}},
    \label{eq:HD}
\end{equation}

\noindent where $R_{+\ell}$ and $R_{-\ell}$ denote the transmitted,
reflected, or scattered intensities under right- and left-handed vortex
beams of magnitude $|\ell|$, respectively.

The experimental relevance of OAM for chiral discrimination was
established by Brullot et al., who first resolved enantiomers via HD
using plasmonic nanoparticle aggregates~\cite{Brullot2016}. Subsequent
work demonstrated giant HD responses exceeding 100\% in purpose-designed
multiscale microstructures exploiting vortical differential scattering
(VDS)~\cite{Ni2021PNAS}, and in inverse-designed chiral geometries
optimized to maximize OAM coupling~\cite{PanNano2025}. Direct numerical
studies have further confirmed that efficient HD coupling requires
spatial overlap between the vortex ring and the chiral
geometry~\cite{Ni2021nano,Reddy2018,Wozniak2019}. Nevertheless, most
experimental demonstrations to date have relied on $\sim$800~nm
femtosecond sources and on plasmonic substrates or expensive
photoresists, creating significant barriers to broader adoption.

Here we address this gap by introducing an all-standard-optics platform
for both the fabrication and the OAM characterization of microscale
chiral polymer structures. On the fabrication side, digital micromirror device (DMD)-based maskless
photolithography combined with capillarity-induced self-assembly
deterministically encodes structural handedness into commercial acrylate
resin micropillars. The same DMD-based optical printing in a similar acrylate 
resin has been successfully used for realising a sensitive and rapid alcohol 
sensor~\cite{Xu2024}. On the characterization side, a continuous-wave
532~nm laser and a liquid-crystal spatial light modulator (LC-SLM) generate vortex beams whose topological charge
is scanned to probe HD. We measure HD signals of $\sim$30\% at the
single-structure level. Finite-difference time-domain (FDTD) simulations on three-dimensional geometries
reconstructed from high-resolution Scanning Electron Microscopy (HR-SEM)  data reproduce the experimental
spectra, providing mechanistic insight and validating the size-matching
picture. The versatility of the platform is further demonstrated on
G-shaped chiral microstructures, a canonical geometry previously studied
only with electron-beam lithography and nonlinear optical
probes~\cite{Valev2009,Valev2010,Valev2012}. Together, these results
establish a practical, low-cost route to OAM-based chirality
characterization that is broadly accessible to the photonics community.

%% ============================================================
\section{Experimental Section}
%% ============================================================

\subsection{Design and Fabrication of Chiral Assemblies}

A capillary-induced self-assembly strategy was used to generate chiral
microstructures from photopolymerized micropillar arrays. Rectangular-cross-section
micropillars are mechanically anisotropic: they bend preferentially along
the minor axis, where capillary forces exceed the critical restoring
force~\cite{Chandra2009}. When arranged in a spiral pattern, this
cooperative bending produces helices with well-defined handedness;
clockwise tilting yields right-handed assemblies (RHS) and
counterclockwise tilting yields left-handed assemblies
(LHS)~\cite{Hu2020}.

\begin{figure}[H]
    \centering
    \includegraphics[width=0.85\textwidth]{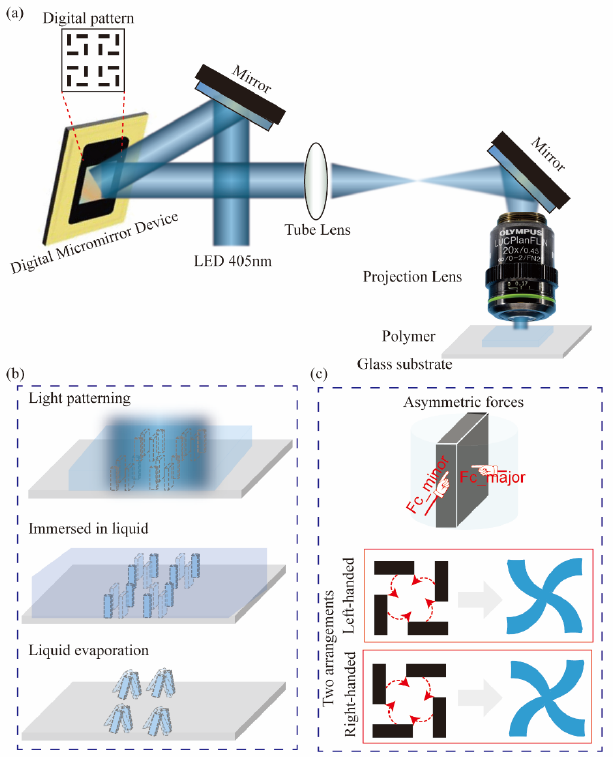}
    \caption{Integration of maskless photolithography with
    capillarity-induced self-assembly. (a) Schematic of the DMD-based
    maskless photopolymerization setup. (b) Post-exposure procedure
    leading to chiral assembly. (c) Schematic illustrating asymmetric
    capillary forces on rectangular micropillars. By prescribing the
    orientation of each rectangular element in the photomask, the
    preferred bending direction is encoded, producing either left- or
    right-handed chiral assemblies.}
    \label{fig:fab}
\end{figure}

Binary rectangular motifs ($7.39\,\mu\mathrm{m} \times
2.46\,\mu\mathrm{m}$) were generated in MATLAB and arranged in
controlled orientations. By rotating each rectangular element relative to
its neighbors, arrays were designed whose preferred bending directions
collectively curve either clockwise or counterclockwise
(Fig.~\ref{fig:fab}c), so that handedness is deterministically encoded
in the projected DMD pattern before any self-assembly occurs.

The fabrication setup (Fig.~\ref{fig:fab}a) employed a 405~nm LED
(Thorlabs M405LP1-C1, 750~mW) as the illumination source. A DMD (Texas
Instruments, $1024\times768$ pixels, 13.68~$\mu$m pixel pitch) spatially
modulated the intensity distribution according to the uploaded binary
patterns. The patterned light was relayed onto the photoresin film
through an Olympus $10\times$ objective (NA~=~0.3), covering an exposure
area of $1.26\times0.95$~mm$^{2}$.

Glass substrates were cleaned with ammonium peroxydisulfate solution and
functionalized with poly(allylamine hydrochloride) (PAH) to promote
adhesion. A 13~$\mu$m film of commercial acrylate photoresin (eSUN
GP001-JC) was deposited by blade-coating. Exposure was performed at
0.71~mW/mm$^{2}$ for 4.5~s to achieve high-aspect-ratio pillars while
minimizing over-polymerization. Samples were developed in 1-propanol for
5~min, rinsed with isopropanol, and dried under ambient conditions.
Capillary forces acting on the anisotropic pillars during drying drove
the directional self-assembly into chiral configurations
(Fig.~\ref{fig:fab}b).

\subsection{Chiroptical Characterization with Vortex Beams}

\begin{figure}[H]
    \centering
    \includegraphics[width=0.9\textwidth]{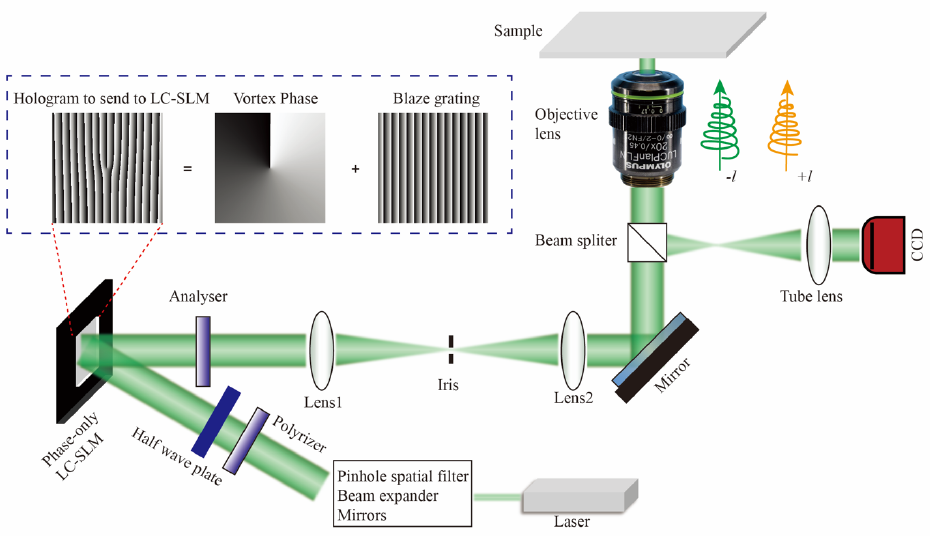}
    \caption{Optical setup for helical dichroism measurements. A
    phase-only spatial light modulator (SLM) encodes vortex phases with
    topological charge $\pm|\ell|$ to generate left- and right-handed
    vortex beams. A 4f relay system filters the first diffraction order
    and sizes the beam to match the microscope objective entrance pupil.}
    \label{fig:setup}
\end{figure}

Vortex beams were generated with a phase-only LC-SLM (Holoeye
HED6010-L-NIII-HR). As shown in Fig.~\ref{fig:setup}, a 532~nm (633~nm) He-Ne
laser (Uniphase 102-3, 4~mW) was expanded and collimated before
incidence on the SLM. A computer-generated helical phase mask, combined
with a blazed grating, was uploaded to the SLM to imprint the azimuthal
phase modulation of Eq.~(\ref{eq:vortex}) while spatially separating
diffraction orders. A polarizer before the SLM and an analyzer after it
ensured the linear polarization state required for efficient phase
modulation. The first diffraction order carrying the vortex beam was
isolated and relayed through a 4f optical system for spatial filtering
and beam resizing, ensuring that it filled the entrance pupil of the
microscope objective. Intensity patterns, with and without the
microstructures, were captured with a CCD camera (BFS-U3-16S2M-CS) and
used to compute HD via Eq.~(\ref{eq:HD}).

%% ============================================================
\section{Results and Discussion}
%% ============================================================

\subsection{Fabricated Chiral Structures}

\begin{figure}[H]
    \centering
    \begin{subfigure}[b]{0.47\textwidth}
        \includegraphics[width=\textwidth]{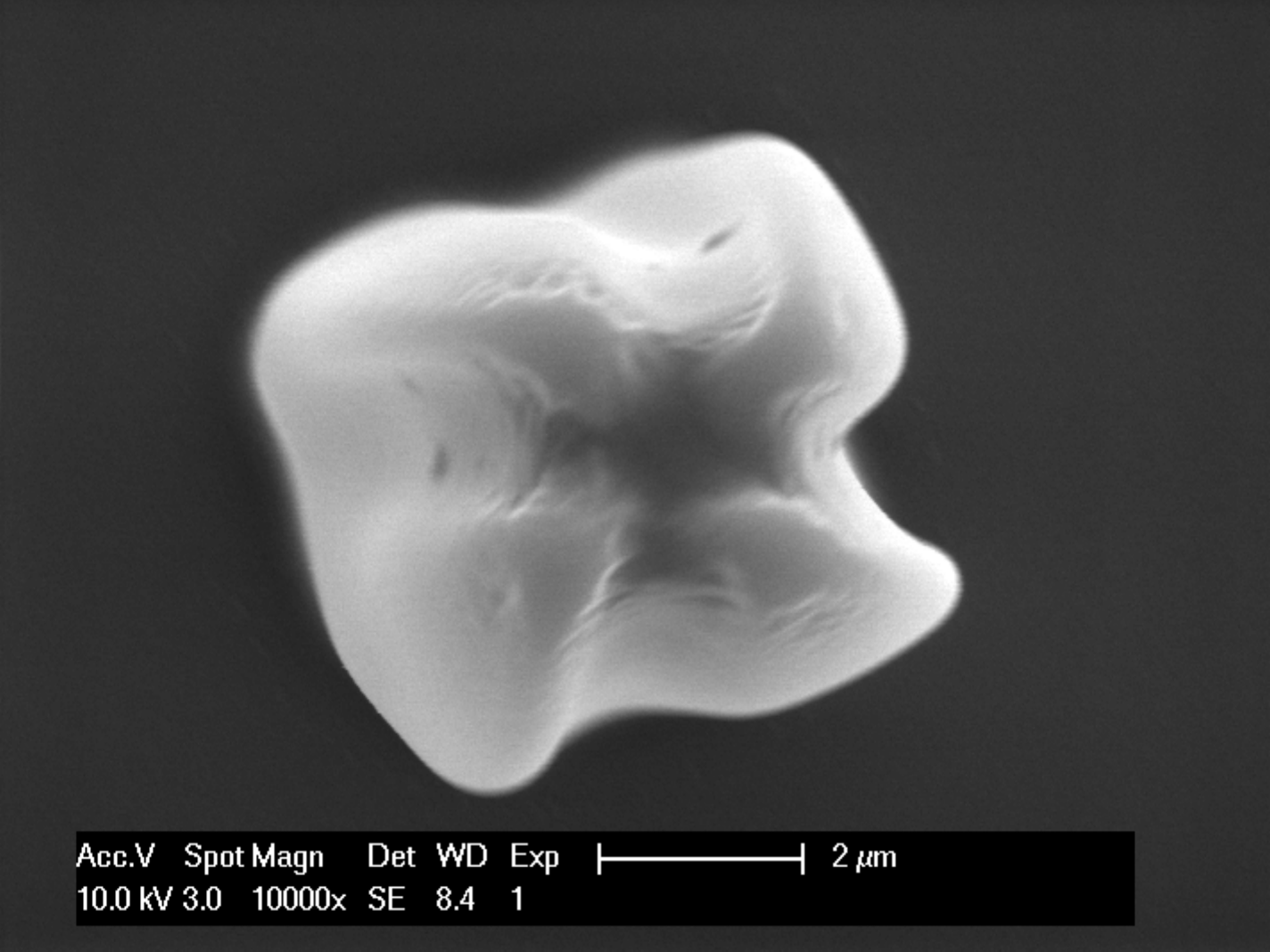}
        \caption{LHS}
    \end{subfigure}
    \hfill
    \begin{subfigure}[b]{0.47\textwidth}
        \includegraphics[width=\textwidth]{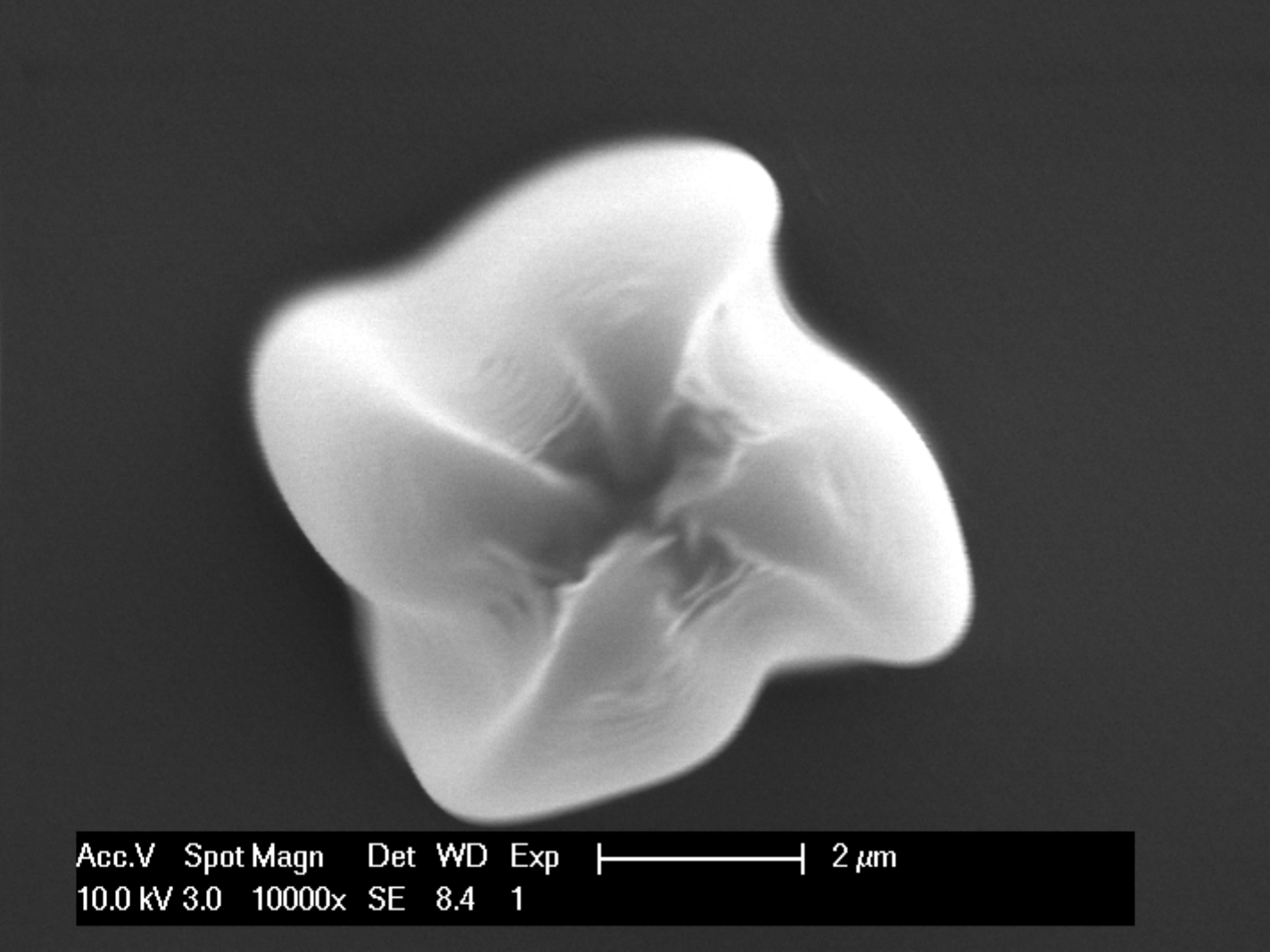}
        \caption{RHS}
    \end{subfigure}
    \caption{Scanning electron microscopy images of fabricated complex
    chiral assemblies. (a) Left-handed structure (LHS). (b) Right-handed
    structure (RHS). The opposite tilt directions of the four pillars
    encode the structural handedness. Scale bar: 2~$\mu$m.}
    \label{fig:SEM}
\end{figure}

\begin{figure}[H]
    \centering
    \includegraphics[width=\textwidth]{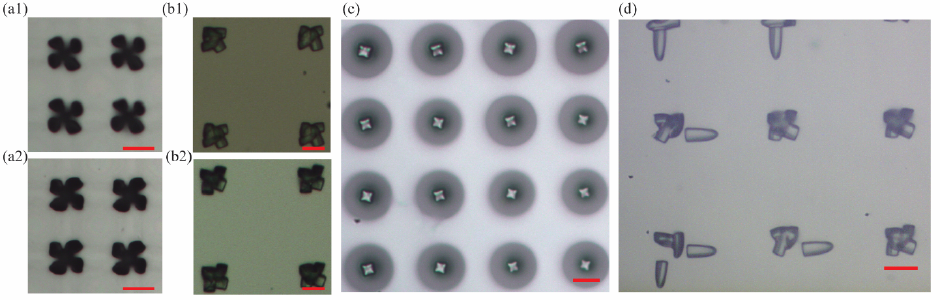}
    \caption{Diversity of assemblies produced by combining DMD-based
    photolithography with capillarity-induced self-assembly.
    (a1,~a2) Left- and right-handed assemblies printed with the 4$\times$
    objective. (b1,~b2) Left- and right-handed assemblies printed with the
    10$\times$ objective. (c) Combined left/right arrays printed with the
    20$\times$ objective. (d) Partially collapsed structures printed with
    the 10$\times$ objective, enabling direct estimation of pillar height.
    Scale bars: (a) 50~$\mu$m; (b) 15~$\mu$m; (c) 10~$\mu$m;
    (d) 15~$\mu$m.}
    \label{fig:arrays}
\end{figure}

The DMD-based fabrication method successfully produced chiral assemblies
in both left- and right-handed configurations. By switching objectives
and modifying the uploaded digital mask, we produced arrays spanning
feature sizes of $54.9\pm0.8$~$\mu$m, $11.2\pm0.5$~$\mu$m, and
$4.6\pm0.3$~$\mu$m without any change to the optical setup or
mechanical settings. Scanning electron microscopy images of the complex
twisted assemblies are shown in Fig.~\ref{fig:SEM}. The opposite tilt
directions of the four intertwined pillars are clearly visible in each
enantiomer. Optical microscopy images of a broader range of fabricated
assemblies, including combined left/right arrays and partially collapsed
structures that allow direct estimation of pillar height
($11.18\pm0.52$~$\mu$m), are provided in Fig.~\ref{fig:arrays}.

Initial CD measurements on these structures using a commercial
spectrometer (Jasco J-810) yielded signals indistinguishable from
background noise (Fig.~\ref{fig:CD} in Supporting Information), confirming the
fundamental insensitivity of SAM-based techniques when structural
dimensions greatly exceed the probing wavelength.

\subsection{Vortex Beam Generation}

\begin{figure}[H]
    \centering
    \includegraphics[width=0.60\textwidth]{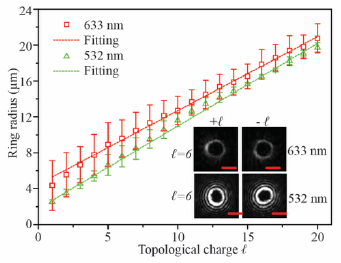}
    \caption{Measured ring radius of generated vortex beams versus
    topological charge $|\ell|$ at two wavelengths. The approximately
    linear scaling allows direct size-matching between the vortex ring
    and the target microstructure.}
    \label{fig:vortex}
\end{figure}

The intensity profiles of the generated vortex beams are shown in
Fig.~\ref{fig:vortex}. The expected donut-shaped distributions are
observed, with ring radius increasing monotonically with $|\ell|$. For
practical implementations with finite modulator aperture, the ring radius
scales approximately linearly with $|\ell|$~\cite{Curtis2003}, consistent
with our measurements at two wavelengths. This linear tunability is key:
it allows the vortex ring to be size-matched to the chiral structure by
simply adjusting the topological charge.

\subsection{Helical Dichroism: Simple Chiral Assemblies}

\begin{figure}[H]
    \centering
    \includegraphics[width=\textwidth]{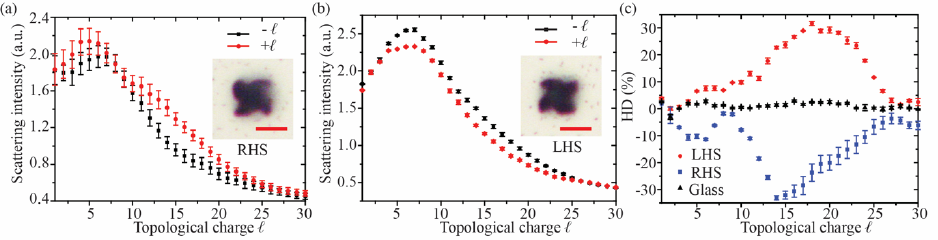}
    \caption{Helical dichroism of simple chiral structures of approx. 15 mm. Scattering
    intensity difference maps for (a) right-handed structures (RHS) and
    (b) left-handed structures (LHS) under excitation with vortex beams
    of topological charge $|\ell| = 1$--25 (insets: optical images, scale
    bar 15~$\mu$m). (c) HD spectra for a single RHS, LHS, and achiral
    control. Error bars represent the standard deviation from three
    independent measurements. The peak HD near $\ell = 14$--18 coincides
    with size-matching between the vortex ring and the $\sim$15~$\mu$m
    structure.}
    \label{fig:HD_simple}
\end{figure}

We first probed the simpler spiral micropillar assemblies, which have
characteristic diameters of $\sim$15~$\mu$m and serve as a controlled
test case. The scattering intensity maps in
Fig.~\ref{fig:HD_simple}(a,b) show that RHS preferentially scatters
vortex beams with $+\ell$, while LHS preferentially scatters beams with
$-\ell$. This enantiomer-specific response produces the approximately
mirror-symmetric HD spectra of Fig.~\ref{fig:HD_simple}(c), with peak
signals of $\sim$30\% near $\ell = 14$--18, precisely where the vortex
ring diameter matches the structural diameter. The achiral control
exhibits a vanishing HD throughout, confirming the chiroptical origin of
the signal.

The slight asymmetry between LHS and RHS spectra---maximum HD at
$\ell \approx 17$ for LHS versus $\ell \approx 14$ for RHS---is
attributed to fabrication variability in pillar height, tilt, and
lateral spacing, as well as to minor lateral offsets between the vortex
beam axis and the structure center during measurement.

\begin{figure}[H]
	\centering
	\includegraphics[width=0.55\textwidth]{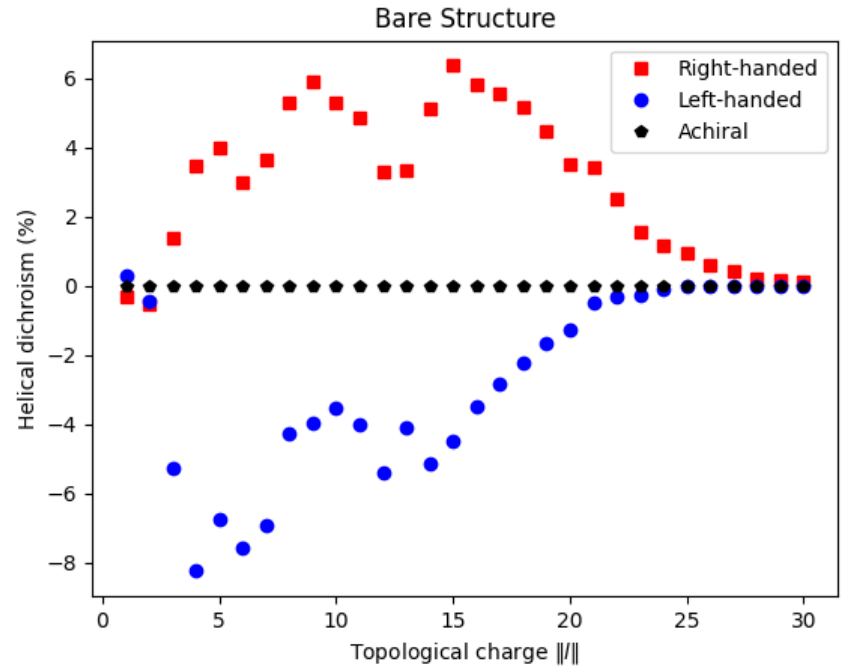}
	\caption{Helical dichroism of simple twisted chiral assemblies 
		obtained from FDTD calculations on
		reconstructed 3D geometries. The near-mirror symmetry between
		enantiomers and the vanishing achiral signal confirm the robustness
		of OAM as a handedness-sensitive probe.}
	\label{fig:HD_simple-sim}
\end{figure}

Numerical (FDTD) simulations of these structures have been realized, 
owing to a reconstruction via HR-SEM images. The reconstructed structures 
are shown in Fig.~\ref{fig:LG} (bottom) both for the RHS and LHS (slightly different) enantiomers. 
Simulations were performed with LG$_{0,\ell}$ beams at 532~nm with beam
waist $\omega_0 = 2.5$~$\mu$m. The computational box lateral size was
$12\omega_0$ with longitudinal extent $h + 2\lambda$, and boundaries
were terminated with 40-layer stable perfectly matched layers (PMLs).
The number of grid points per wavelength was set to at least 10. The
simulated HD spectra (Fig.~\ref{fig:HD_simple-sim}) reproduce the key
experimental features: opposite signs for the two enantiomers, peak
magnitudes of 24--32\% (the simulations have been performed at 
half-scale of the structures, leading to a factor 4 in the simulated HDs), 
and a near-zero achiral response. As such, the agreement between experiments 
and simulations is noteworthy very good. 

\subsection{Helical Dichroism: Complex Twisted Assemblies and FDTD
Simulations}

\begin{figure}[H]
    \centering
    \begin{subfigure}[b]{0.47\textwidth}
        \includegraphics[width=\textwidth]{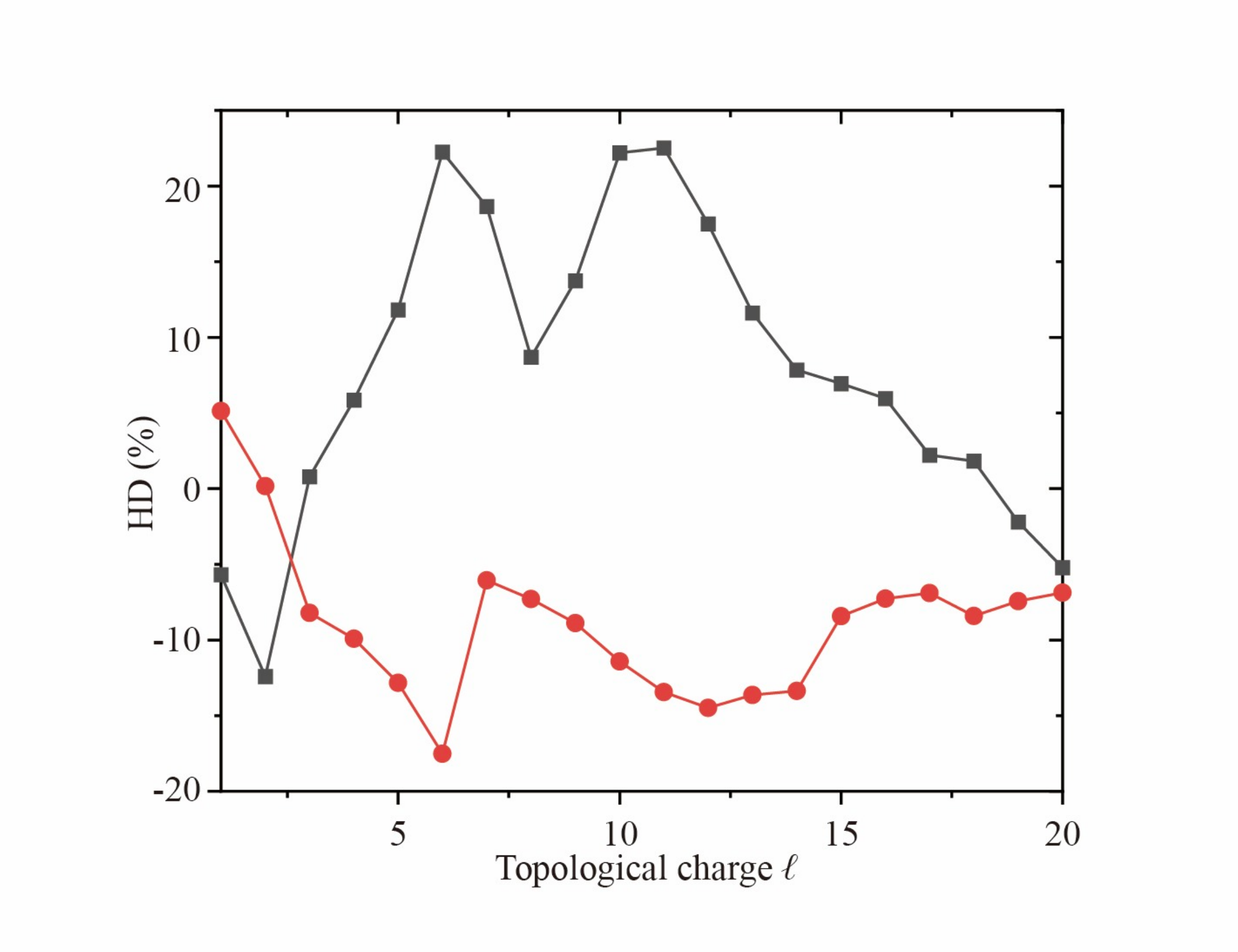}
        \caption{Experimental HD}
    \end{subfigure}
    \hfill
    \begin{subfigure}[b]{0.47\textwidth}
        \includegraphics[width=\textwidth]{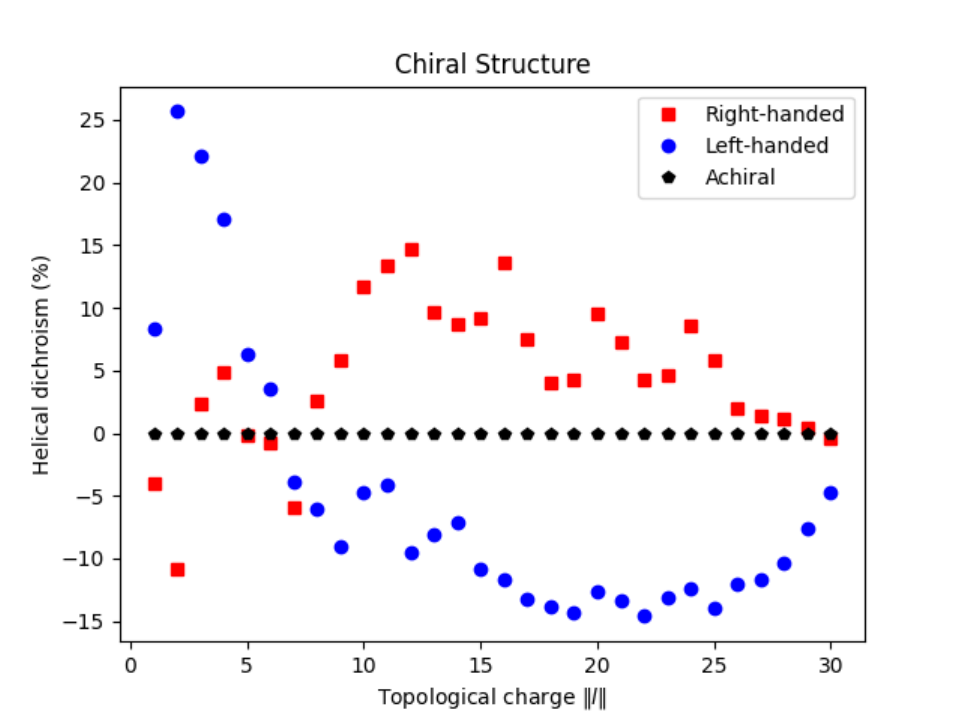}
        \caption{Simulated HD}
    \end{subfigure}
    \caption{Helical dichroism of complex twisted chiral assemblies.
    (a) Experimental HD spectra for LHS (red circles) and RHS (black
    squares) as a function of topological charge $\ell$. (b) Simulated HD
    spectra for RHS (red squares), LHS (blue circles), and an achiral
    reference (black diamonds), obtained from FDTD calculations on
    reconstructed 3D geometries. The near-mirror symmetry between
    enantiomers and the vanishing achiral signal confirm the robustness
    of OAM as a handedness-sensitive probe.}
    \label{fig:HD_complex}
\end{figure}

To access more complex chiroptical behavior and to validate our
simulation framework against real three-dimensional geometries, we
extended the study to the more strongly twisted assemblies. Their
experimental HD spectra are shown in Fig.~\ref{fig:HD_complex}(a).
Both enantiomers exhibit clear, opposite-signed HD signals across the
entire topological charge range probed, with magnitudes reaching
$\sim$20--25\%, again demonstrating high chiroptical contrast from
an all-polymer, all-standard-optics system.

\begin{figure}[H]
    \centering
    \begin{subfigure}[b]{0.47\textwidth}
        \includegraphics[width=\textwidth]{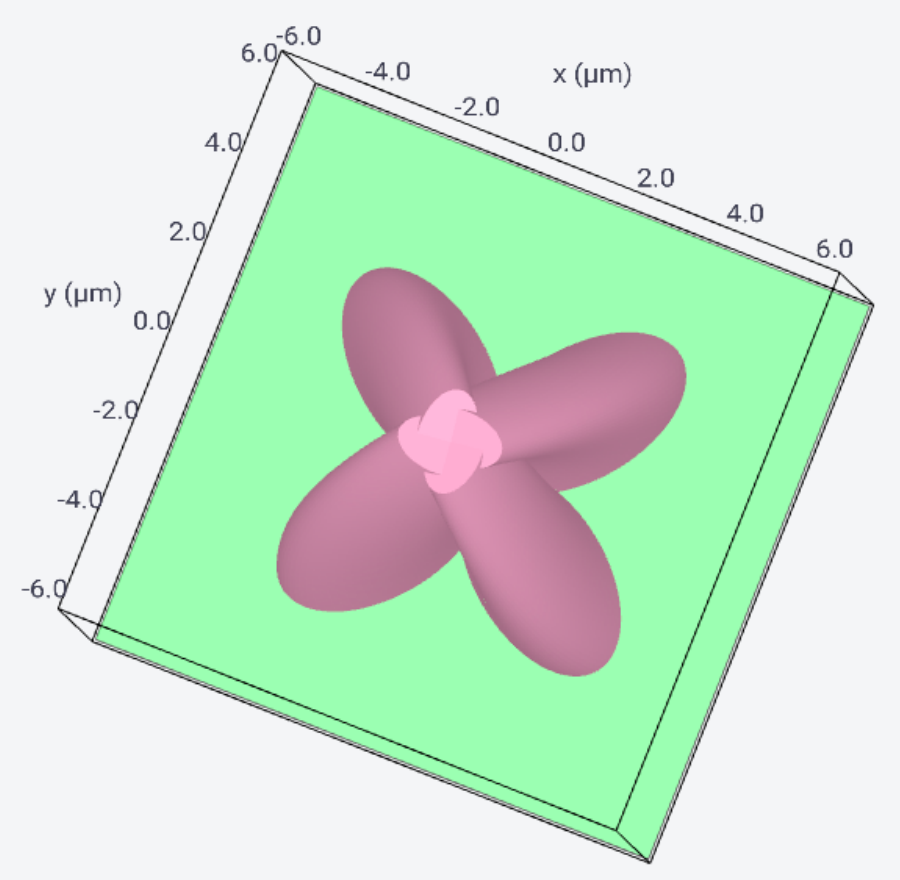}
        \caption{LHS}
    \end{subfigure}
    \hfill
    \begin{subfigure}[b]{0.47\textwidth}
        \includegraphics[width=\textwidth]{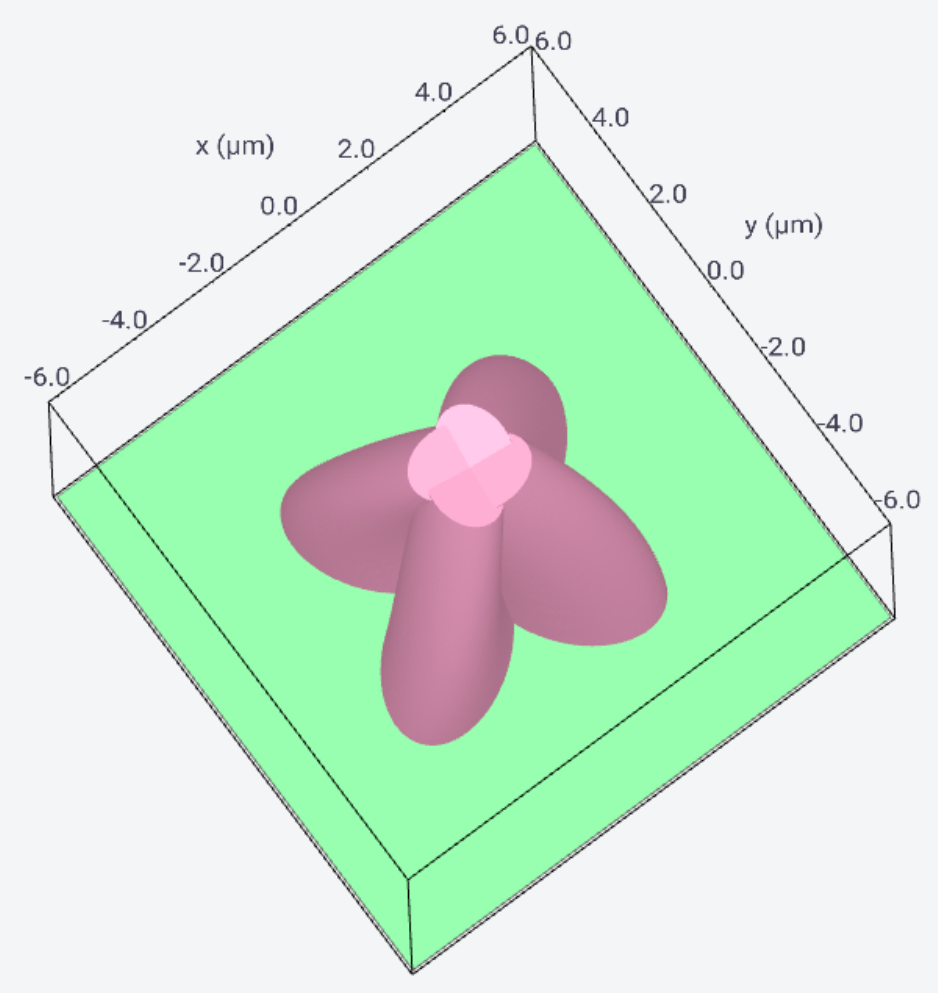}
        \caption{RHS}
    \end{subfigure}
    \caption{Three-dimensional geometries of complex chiral assemblies
    reconstructed from (HR) Scanning Electron Microscopy  data and used as FDTD
    simulation inputs. (a) Left-handed structure (LHS). (b) Right-handed
    structure (RHS). Dimensions in $\mu$m.}
    \label{fig:3D}
\end{figure}

\begin{figure}[H]
    \centering
    \begin{subfigure}[b]{0.47\textwidth}
        \includegraphics[width=\textwidth]{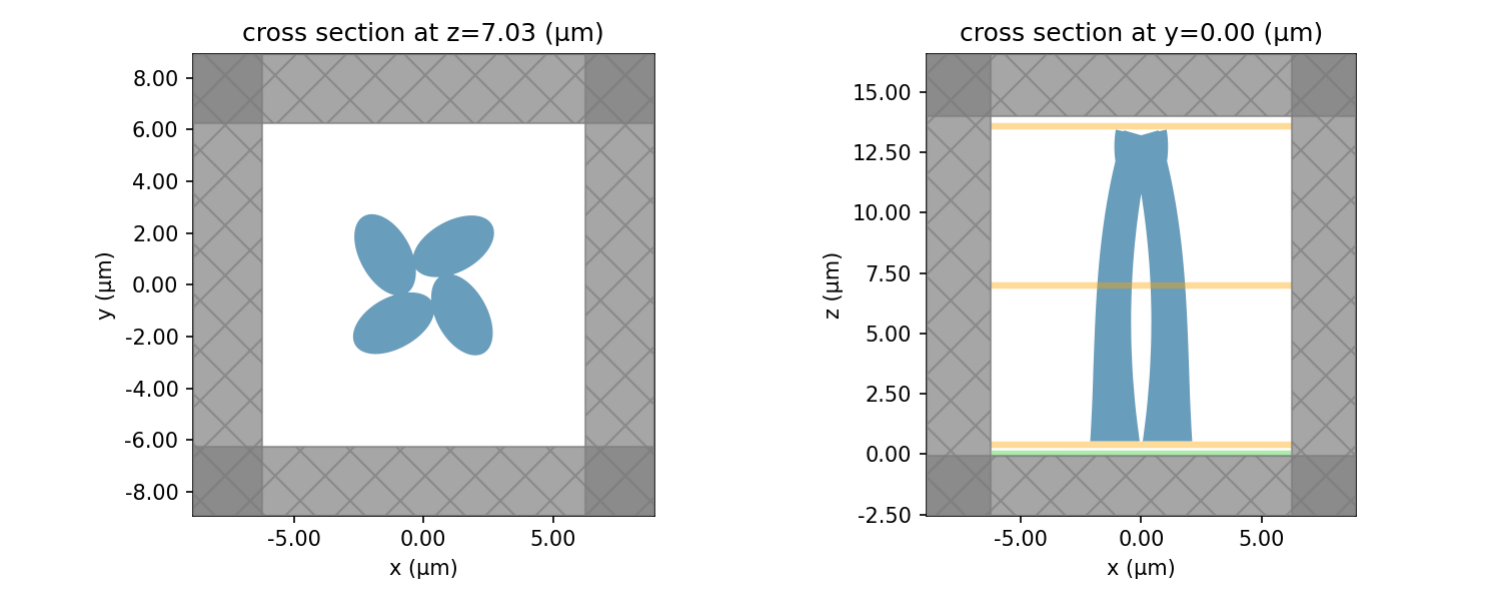}
        \caption{LHS cross-sections}
    \end{subfigure}
    \hfill
    \begin{subfigure}[b]{0.47\textwidth}
        \includegraphics[width=\textwidth]{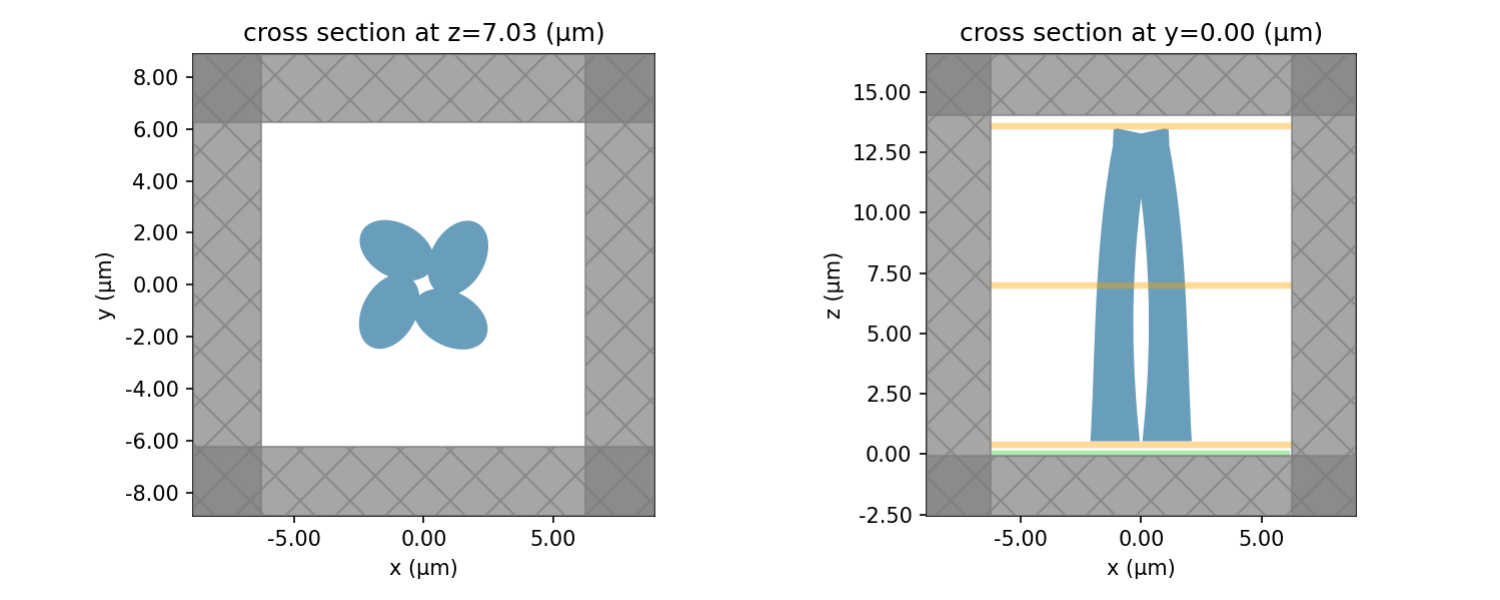}
        \caption{RHS cross-sections}
    \end{subfigure}
    \caption{Simulation geometry cross-sections. Left panel: horizontal
    cross-section at $z = 7.03$~$\mu$m showing the four-lobe transverse
    profile. Right panel: vertical cross-section at $y = 0$~$\mu$m
    showing pillar height and the positions of the source plane (green)
    and detector planes (yellow); the central detector is used for HD
    computation. (a) LHS. (b) RHS.}
    \label{fig:simgeo}
\end{figure}

To obtain quantitative agreement with these measurements, we
reconstructed full three-dimensional geometries from (HR) Scanning
 Electron Microscopy data. Representative three-dimensional
renderings and their cross-sectional views are shown in
Figs.~\ref{fig:3D} and~\ref{fig:simgeo}, respectively. The four-lobe
transverse profile visible in the $z = 7.03$~$\mu$m cross-section
reflects the four-pillar motif of the chiral assembly, and the
side-view confirms the full pillar extent before twisting. These
reconstructed geometries were used directly as inputs for FDTD
calculations, without simplifying assumptions about the structural shape.

Simulations were performed in a similar way to the simple structures here above. The
simulated HD spectra (Fig.~\ref{fig:HD_complex}b) reproduce the key
experimental features: opposite signs for the two enantiomers, peak
magnitudes of 15--25\%, and a near-zero achiral response. The residual
quantitative discrepancy is attributed to two sources. First, minor
lateral misalignment between the vortex beam axis and the structure
center during measurement, since geometric overlap strongly influences
HD amplitude. Second, imperfections in the volumetric reconstruction
arising a detailed evaluation of intrinsically 2D SEM micrographs, 
which can introduce errors in the 3D reconstructed geometry.

\subsection{Generality of the Platform: G-Shaped Chiral Structures}

To demonstrate that the platform is not limited to spiral pillar
assemblies, we fabricated G-shaped chiral microstructures by simply
modifying the DMD pattern---no change to the optical setup or processing
protocol was required. The G-shape geometry was originally introduced by
Valev et al.\ and studied by electron-beam lithography and nonlinear
optical spectroscopy~\cite{Valev2009,Valev2010,Valev2012}. Our approach
reproduces this canonical chiral geometry at an enlarged scale and
characterizes it directly by HD, providing a more accessible route to
study such designs.

\begin{figure}[H]
    \centering
    \includegraphics[width=0.55\textwidth]{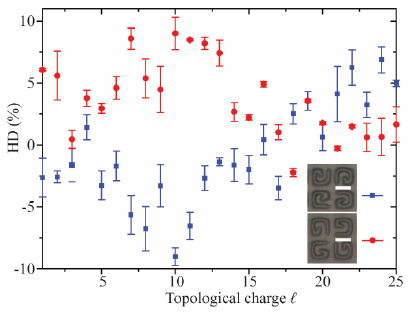}
    \caption{HD spectra for a G-shaped chiral structure and its mirror
    enantiomer (scale bar: 10~$\mu$m). Error bars represent the standard
    deviation from three independent measurements. The opposite-signed
    spectra confirm handedness sensitivity for a geometry unrelated to
    the spiral pillar motif, demonstrating the generality of the OAM
    characterization approach.}
    \label{fig:Gshape}
\end{figure}

As shown in Fig.~\ref{fig:Gshape}, opposite-signed HD spectra are
clearly resolved for the two G-shaped enantiomers, confirming that the
OAM characterization approach is geometry-independent. This flexibility
highlights a key advantage of the platform: a single, programmable
DMD-based fabrication step can produce virtually any two-dimensional
chiral motif, which can then be characterized immediately using the same
SLM-based setup.

%% ============================================================
\section{Conclusion}
%% ============================================================

We have demonstrated a simple, cost-effective platform for fabricating
and characterizing microscale chiral polymer structures using photonic
orbital angular momentum. DMD-based maskless photolithography with
capillarity-induced self-assembly deterministically encodes structural
handedness into commercial acrylate resin micropillars, while a standard
CW laser combined with an LC-SLM generates tunable vortex beams for HD
measurements. HD signals of $\sim$30\% were measured at 532~nm for
$\sim$15~$\mu$m structures, without recourse to femtosecond sources,
plasmonic substrates, or costly photoresists. FDTD simulations based on
three-dimensional geometries reconstructed from (HR) Scanning Electron Microscopy  data
reproduce the experimental spectra and confirm the size-matching
mechanism: peak HD occurs when the vortex ring diameter matches the
structural diameter. Near-mirror-symmetric HD spectra for opposite-handed
enantiomers, and a vanishing signal for achiral controls, establish OAM
as a robust, spatially selective chiroptical probe at the single-structure
level. Extension to G-shaped chiral microstructures---fabricated by a
simple mask change---demonstrates the generality of the approach.

Beyond the specific systems studied here, this work establishes that
high-contrast chiroptical characterization at the microscale is
achievable with instrumentation standard in an optics laboratory. This
accessibility, combined with the programmability of both the fabrication
(DMD patterns) and the probe (SLM topological charge), opens practical
routes toward OAM-driven chiral sensing, enantioselective chemical
detection, photonic circuitry, and optical manipulation. The findings
further contribute to the growing body of chiroptical research by showing
that material structuring combined with structured light can drive the
development of next-generation photonic and optoelectronic devices.

%% ============================================================
\section*{Author Contributions}
%% ============================================================

J.X.\ built the setup, performed the experiments, and wrote the original manuscript; R.S.\ developed the software for data analysis; Y.dC.\ provided ideas and guided the work, R.V.\ conducted the simulations, provided conceptual ideas, added the simulation part to the manuscript, K.C.\ supervised the project, all authors contributed to revising the manuscript.  

%% ============================================================
\section*{Acknowledgements}
%% ============================================================

The authors gratefully acknowledge internal funding (BOF) by the University of Leuven Research Council through the C16/23/004 project "Exploring the interaction of spin and orbital angular momentum of light with chiral photonic structures".

%% ============================================================
\section*{Conflict of Interest}
%% ============================================================

The authors declare no conflict of interest.

%% ============================================================
\section*{Data Availability Statement}
%% ============================================================

The data that support the findings of this study are available from the
corresponding author upon reasonable request.

%% ============================================================
%% SUPPORTING INFORMATION
%% ============================================================
\clearpage

%% Switch to "S"-prefixed numbering for sections, figures, equations, tables
\setcounter{section}{0}
\setcounter{figure}{0}
\setcounter{equation}{0}
\setcounter{table}{0}
\renewcommand{\thesection}{S\arabic{section}}
\renewcommand{\thesubsection}{S\arabic{section}.\arabic{subsection}}
\renewcommand{\thefigure}{S\arabic{figure}}
\renewcommand{\thetable}{S\arabic{table}}
\renewcommand{\theequation}{S\arabic{equation}}

\begin{center}
{\Large\textbf{Supporting Information}}
\end{center}

\vspace{1em}

%% ============================================================
\section{CD Spectrum: Evidence for the Failure of SAM-Based Detection}
%% ============================================================

To establish the need for an OAM-based approach, we first attempted to
characterize the fabricated chiral structures with a conventional
circular dichroism (CD) spectrometer (Jasco J-810). CD relies on photonic
spin angular momentum (SAM) and its signal strength depends on the
interference between electric-dipole (E1) and magnetic-dipole (M1)
transitions. When the structural feature size greatly exceeds the probing
wavelength, this interference is suppressed and the CD signal vanishes.

\begin{figure}[H]
    \centering
    \includegraphics[width=0.55\textwidth]{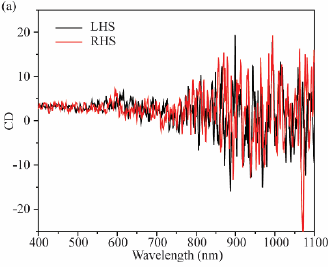}
    \caption{CD spectra of chiral microstructures with a characteristic
    size of 5~$\mu$m. The signal is indistinguishable from noise across
    the entire accessible spectral range, confirming the fundamental
    insensitivity of SAM-based techniques for microscale chiral objects.}
    \label{fig:CD}
\end{figure}

As shown in Fig.~\ref{fig:CD}, measurements on structures with a
characteristic size of 5~$\mu$m yielded signals entirely indistinguishable
from the background noise across the accessible spectral range
(400--1100~nm). This null result directly motivates the use of photonic
OAM, whose transverse ring diameter can be tuned to match the structural
scale, restoring strong light--matter coupling.

%% ============================================================
\section{Chemicals and Fabrication Protocol}
%% ============================================================

\subsection{Chemicals}

A commercial acrylate-based photoresin (eSUN Standard Resin GP001-JC) was
purchased and used as the recording medium for the generation of chiral
assemblies. Ammonium peroxydisulfate acidic powder, sulfuric acid (95\%),
poly(allylamine hydrochloride) (PAH), and isopropanol (99+\%) were
purchased from Sigma-Aldrich and used as received.

\subsection{Substrate Preparation}

Glass substrates were cleaned by immersion in ammonium peroxydisulfate
acidic solution to remove organic contaminants. To improve adhesion
between the polymer structures and the substrate, the cleaned glass
surface was subsequently treated with poly(allylamine hydrochloride)
(PAH) by spin-coating a 1~mg/mL aqueous PAH solution and allowing it to
dry under ambient conditions.

\subsection{Photoresin Film Deposition}

A relatively thick film of the photoresin (13~$\mu$m) was deposited onto
the treated glass substrate by blade-coating. Film thickness was
controlled by the blade gap and the resin viscosity, and was verified by
profilometry on reference samples. The film thickness determines the
micropillar height after exposure and development, and therefore controls
the aspect ratio of the structures and the capillary-force-driven
self-assembly behavior.

\subsection{Exposure and Development}

Exposure intensity and time were set at 0.71~mW/mm$^{2}$ and 4.5~s,
respectively, to achieve high-aspect-ratio micropillars while minimizing
over-polymerization that would merge neighboring pillars and suppress
self-assembly. Following exposure, samples were developed in 1-propanol
for 5~min to remove unreacted and weakly crosslinked resin. Samples were
then rinsed with isopropanol and dried under ambient conditions. During
drying, capillary forces acting on the anisotropic rectangular pillars
drove the unidirectional bending and self-assembly into chiral
configurations.

%% ============================================================
\section{Laguerre--Gaussian Beam Profiles Used in Simulations}
%% ============================================================

\begin{figure}[H]
    \centering
    \includegraphics[width=0.85\textwidth]{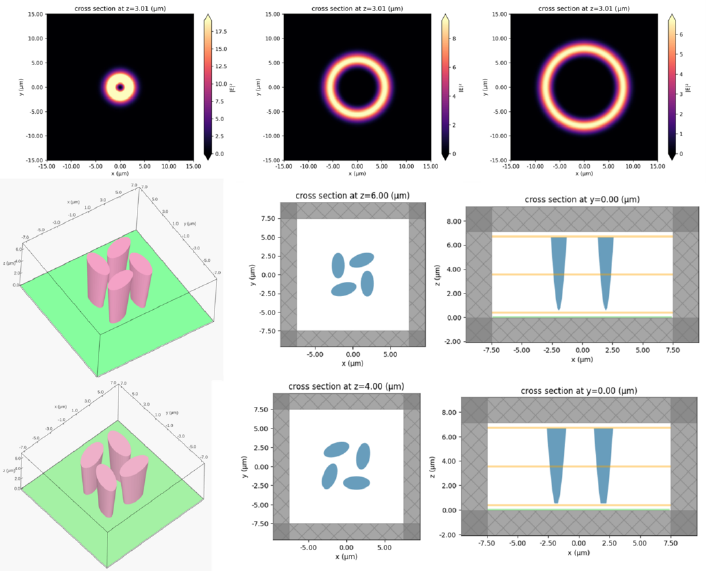}
    \caption{Top line: Laguerre--Gaussian beam intensity profiles used in FDTD
    simulations, for beam waist $\omega_0 = 2.5$~$\mu$m and
    $\lambda = 532$~nm. LG$_{0,\ell}$ modes with varying topological
    charge $\ell$ are shown. The ring radius scales with $|\ell|$,
    consistent with experimental measurements. Middle line: 3D rendering and simulation geometry 
    cross-sections of horizontal cross-section for the RHS simple structure
    at $z = 6.00$~$\mu$m showing the four frustra transverse
    profile and vertical cross-section at $y = 0$~$\mu$m
    showing pillar height and the positions of the source plane (green)
    and detector planes (yellow); the central detector is used for HD
    computation. Bottom line, idem for the LHS simple structure.}
    \label{fig:LG}
\end{figure}

FDTD simulations were performed using LG$_{0,\ell}$ (Laguerre--Gaussian)
beams as the excitation source. Figure~\ref{fig:LG} shows the computed
intensity profiles for the beam waist $\omega_0 = 2.5$~$\mu$m at
$\lambda = 532$~nm and varying topological charge $\ell$. The donut
radius increases with $|\ell|$, in agreement with the experimental
observations reported in Fig.~\ref{fig:vortex} of the main manuscript. The $p = 0$
radial index ensures a single-ring intensity profile with no subsidiary
rings, which simplifies the spatial overlap integral with the chiral
structure.

Figure~\ref{fig:LG} (bottom) shows a 3D representative reconstruction of both
the RHS and LHS simple structures, together with their cross-sectional areas in 
different planes, highlighting their opposite handedness, as reasonably good enantiomers. 
Their simulated helical dichroism is shown in Fig.~\ref{fig:HD_simple-sim} of the main manuscript.

%% ============================================================
\section{Simulated intensity profiles of simple structures}
%% ============================================================

\begin{figure}[H]
	\centering
	\includegraphics[width=0.85\textwidth]{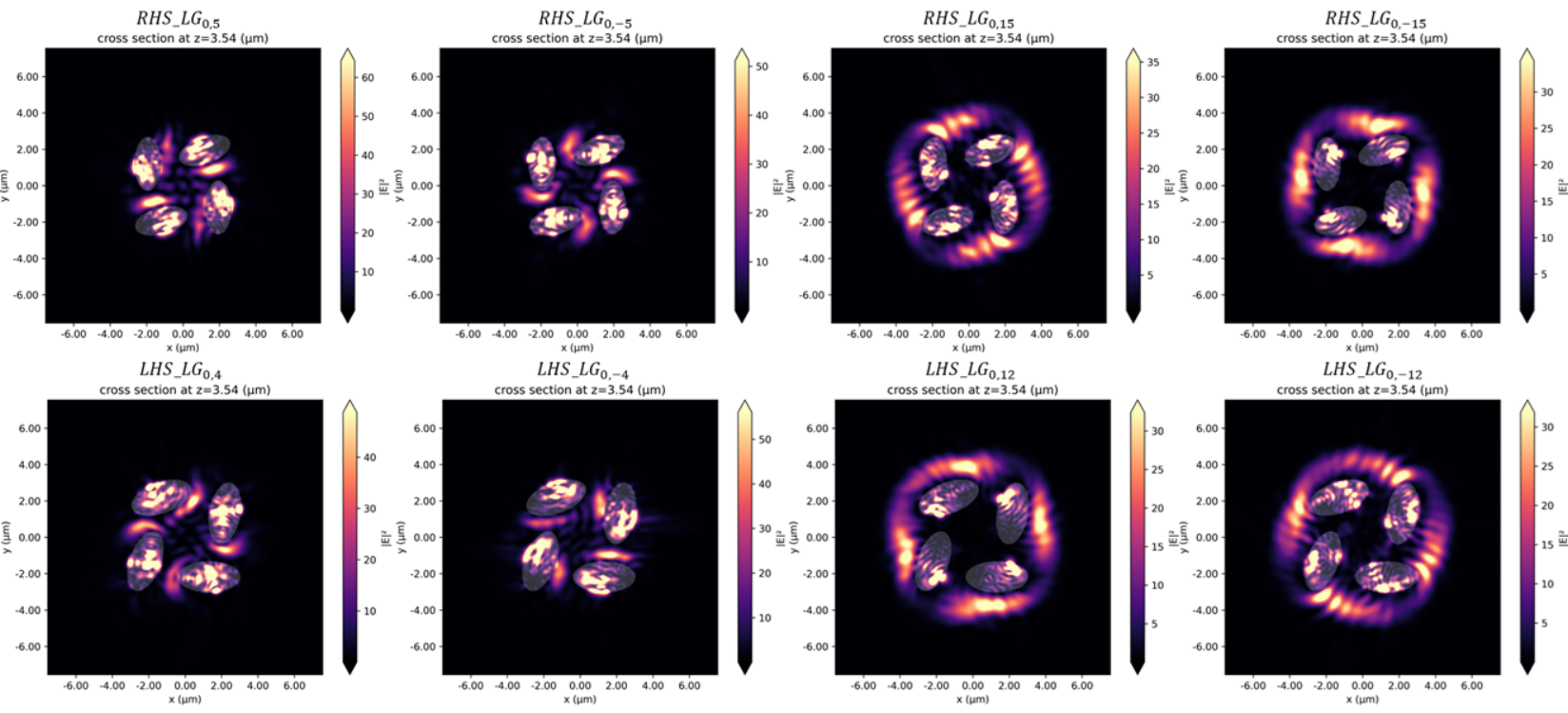}
	\caption{FDTD-simulated intensity distributions under Laguerre--Gaussian excitation. Top row: right-handed structure (RHS); bottom row: left-handed structure (LHS). The structures are illuminated with LG$_{0,\ell}$ beams of opposite orbital angular momentum (OAM) signs ($\pm \ell$) at $\lambda = 532$ nm and beam waist $\omega_0 = 2.5\,\mu$m. Cross-sectional intensity maps (at fixed $z$) reveal a pronounced handedness-dependent response. For the RHS, excitation with one OAM sign induces a well-defined clockwise spiral of the internal field, whereas the opposite sign produces a non-spiral or discontinuous ring distribution. In contrast, the LHS exhibits the reversed behavior, with a counter-clockwise spiral forming under the opposite OAM sign and a disrupted ring for the corresponding counterpart. These asymmetric field distributions highlight the enantiomer-specific interaction between structured light and chiral geometry, consistent with the helical dichroism response reported in Fig.~\ref{fig:HD_simple-sim} of the main manuscript.}
	\label{fig:SIP}
\end{figure}

The simulated intensity profiles reveal a clear dependence on both the handedness of the structure and the sign of the orbital angular momentum of the incident LG$_{0,\ell}$ beams. For the right-handed structure (RHS), one sign of $\ell$ induces a pronounced clockwise spiral of the internal field distribution, while the opposite sign results in a non-spiral or partially disrupted ring. In particular, the field along the ring remains continuous for one OAM sign but becomes fragmented for the other. This asymmetry directly contributes to the strong features observed in the helical dichroism spectra. Conversely, the left-handed structure (LHS) exhibits the opposite behavior: a well-defined counter-clockwise spiral emerges for the corresponding OAM sign, whereas the opposite excitation leads to a disrupted or discontinuous intensity distribution. This inversion of the field topology between RHS and LHS confirms the enantiomeric nature of the structures and demonstrates their selective coupling to the handedness of the optical excitation.

%% ============================================================
\bibliographystyle{unsrtnat}
\bibliography{references}
%% ============================================================

\end{document}